\documentclass[aps,prl,twocolumn,superscriptaddress, 10pt]{revtex4-2}
\usepackage[utf8]{inputenc}
\usepackage[T1]{fontenc}
\usepackage{amsmath,amssymb}
\usepackage{times}
\usepackage{xcolor}
\usepackage{physics}
\usepackage[normalem]{ulem}
\usepackage{xspace}
\usepackage{float}
\usepackage{comment}
\usepackage[colorlinks=true,
            linkcolor=blue,
            urlcolor=cyan,
            citecolor=cyan,
            filecolor=magenta]{hyperref}
\usepackage{mathtools}
\usepackage{graphicx}
\usepackage{dcolumn}
\usepackage{bm}

\usepackage[normalem]{ulem}

\begin{document}



\title{Long range to short range crossover in one dimension}
\author{Mrinal Sarkar}
\affiliation{Institut für Theoretische Physik, Universität Heidelberg, 69120 Heidelberg, Germany 
}
\author{Nicolò Defenu}%
\affiliation{Institut für Theoretische Physik, ETH Zürich, Wolfgang-Pauli-Str.27, 8093 Zürich, Switzerland
}%
\author{Tilman Enss}%
\affiliation{Institut für Theoretische Physik, Universität Heidelberg, 69120 Heidelberg, Germany 
}%

\date{\today}
\begin{abstract}
This work investigates the critical behavior of one-dimensional systems with long-range (LR) interactions, focusing on the crossover to short-range (SR) universality. Through large-scale Monte Carlo simulations of self-avoiding Lévy flights on a 1D lattice, we compute the anomalous dimension $\eta$, the correlation length exponent $\nu$, and the susceptibility exponent $\gamma$ across a wide range of LR decay parameters $\sigma$. Our results provide strong numerical evidence that supports Sak’s scenario. They identify the crossover at $\sigma^* = 1$ and demonstrate the continuity of critical exponents across this point, with strong corrections to scaling. The study also reveals deviations from Flory-type scaling predictions and discusses the limitations of  effective dimension approaches in general. These findings clarify the nature of the LR--SR crossover in low-dimensional systems and open avenues for exploring criticality in disordered and complex networks.
\end{abstract}

\maketitle

Systems with long-range (LR) interactions, characterized by couplings decaying with distance as a power law, have long been a cornerstone in statistical and condensed matter physics\,\cite{campa2009statistical, defenu2023long}. The field has recently experienced a surge of interest due to the possibility to realize synthetic long-range interactions in atomic, molecular, and optical systems promising several practical applications in quantum technology\,\cite{haffner2008quantum, saffman2010quantum, ritsch2013cold, bernien2017probing, monroe2021programmable}. Indeed, LR systems exhibit a variety of rich physical phenomena both in and out of equilibrium, such as ensemble inequivalence, slow relaxation and quasi-stationary states\,\cite{campa2009statistical,defenu2024out}, which evade the fundamental limitations of their short-range (SR) counterparts. 

Sufficiently slowly decaying LR interactions can induce spontaneous symmetry breaking in low-dimensional systems, which is forbidden by the Mermin-Wagner theorem in the local case\,\cite{dyson1969existence, thouless1969long, kosterlitz1976phase}. This property, which was recently verified in several experiments\,\cite{chen2023continuous,feng2023continuous}, highlights the fundamentally different nature of criticality in LR systems.

To develop an understanding of universality and critical phenomena, consider the celebrated ${O}(n)$ model with LR interactions $\sim r^{-(d+ \sigma)}$, with $r$ the Euclidean distance on the lattice, $d$ the underlying lattice dimension and $\sigma$ the LR decay exponent. The model shows three different universality classes depending on the LR parameter $\sigma>0$:
(i) for $ 0 < \sigma \leq d/2$, it belongs to the mean field (MF) universality class, (ii) for $d/2 < \sigma \leq \sigma^{*} $, it falls within the LR university class, and (iii) for $\sigma > \sigma^{*} $, it crosses over to the SR universality class\,\cite{defenu2020criticality}.

\begin{figure}[tbp!]
\centering
\includegraphics[width=0.9\linewidth]{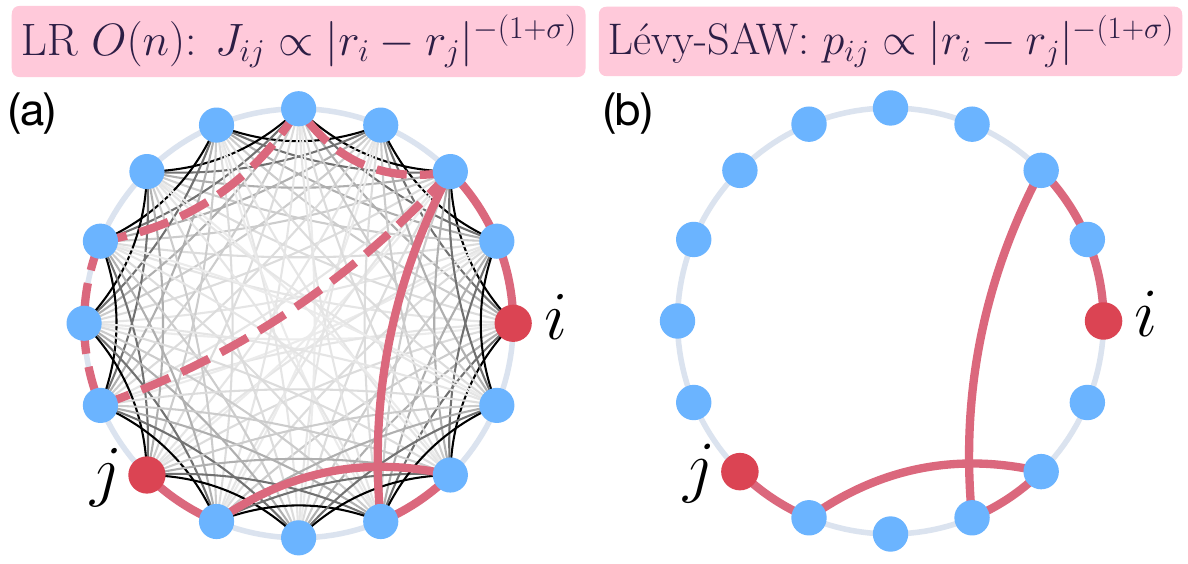}
    \caption{{\bf (a)} Graph representation of typical spin configurations that contribute to the two-point correlation function in the high-temperature expansion of the LR $O(n)$ model. {\bf (b)} Self-avoiding L\'evy flight (L\'evy-SAW) on a $1$D lattice with $L=16$ sites (in our numerics $L\to\infty$), where the interaction (jump-length probability) $\propto r^{-(1+\sigma)}$. As $n \to 0$, a loop contribution (dashed red) in (a) is suppressed, leaving only self-avoiding paths (continuous red), which map to the L\'evy-SAW.}
    \label{fig:On_SAW_mapping}
\end{figure}

While the boundary between MF and LR classes is well understood, the crossover from LR to SR universality classes is less clear. An intuitive approach was presented in the seminal work by Fisher \textit{et al.}~\cite{fisher1972critical}, who by second-order $\epsilon$ expansion located the LR-SR boundary at $\sigma^{*} = 2$ and determined the anomalous dimension $\eta = 2 - \sigma$ in the LR regime. For $\sigma > 2$, SR exponents apply for all $d$. This suggests a discontinuity in $\eta$ at $\sigma^{*}$, jumping from 0 (LR side) to $\eta>0$ (SR side). Subsequently, Sak\,\cite{sak1973recursion} resolved this discontinuity by including higher-order terms in the Renormalization Group (RG) calculations and predicted $\eta$ to be continuous: $\eta = 2 - \sigma$ holds only for $\sigma < \sigma^{*}$ with a shifted boundary $ \sigma^{*} = 2 - \eta_{\rm{SR}}$, where $\eta = \eta_{\rm{SR}}$ reaches the anomalous dimension in the SR limit.

However, Sak's LR–SR crossover scenario has remained under critical scrutiny over the years. Van Enter\,\cite{van1982instability} challenged it by showing that for continuous symmetry $n > 2$, the phase diagram becomes unstable as LR perturbations remain relevant in $2 - \eta_{\rm{SR}} = \sigma^* \leq \sigma \leq 2$. Other field-theoretic studies have presented conflicting views. For instance, Refs.\,\cite{yamazaki1977critical, yamazaki1978comments} argue for the absence of any kink at $\eta = 2 - \eta_{\rm{SR}}$, while Ref.\,\cite{gusmao1983validity} supports Fisher's picture. However, more recent numerical and theoretical investigations have extensively tested Sak's scenario in various symmetry classes and spatial dimensions. For the LR Ising model ($n=1$), both MC simulations in two dimensions (2D)\,\cite{luijten2002boundary, angelini2014relations} and RG approaches\,\cite{brezin2014crossover, defenu2015fixed} support Sak’s picture. However, Refs.\,\cite{picco2012critical,blanchard2013influence} predict the crossover at $\sigma^* = 2$, but this finding has been attributed to the difficulty in capturing logarithmic correction close to the boundary\,\cite{angelini2014relations}. Sak's picture was also confirmed by conformal bootstrap both in $d=2$\,\cite{behan2017long, behan2017scaling} and, very recently, in $d=1$, where $\sigma^{*}=1$\,\cite{benedetti2025one}. For infinite order transitions in the 2D LR XY model ($n=2$), Ref.\,\cite{giachetti2021berezinskii} agrees with Sak’s criterion, while a recent study\,\cite{xiao2024two} suggests that LR behavior may persist up to $\sigma = 2$, favoring Fisher's picture.

The nature of transition symmetry plays a crucial role in the physics of crossover, but studies on discrete symmetries beyond the Ising ($n=1$) case remain limited, with the notable exception of percolation\,\cite{gori2017one}. Although the $n \to 0$ limit belongs to discrete symmetry models, it exhibits qualitatively distinct behavior from the Ising model. Strikingly enough, the $1$D SR Ising model does not exhibit a finite-temperature phase transition, while the $n \to 0$ limit is critical in any dimension\,\cite{de1979scaling, baxter2016exactly,  slade2019self}. Despite this intriguing fact, the critical behavior of LR $O(n \to 0)$ model remains unexplored. This work aims to fill this gap by investigating the LR–SR crossover in the $n \to 0$ limit in $d = 1$.

As first pointed out by de Gennes~\cite{de1972exponents}, the $O(n)$ model in the $n \to 0$ limit is equivalent to the self-avoiding random walk (SAW). In the lattice formulation, the $O(n)$ model is defined by $n$-component spins $\vec{S}_i$ at lattice sites $i$ and normalized as $S_i^2 = n$, with Hamiltonian $\mathcal{H} = - J \sum_{\langle i, j \rangle} \vec{S}_i \cdot \vec{S}_j,$ summed over nearest-neighbor pairs. The central idea is that the two-point correlation function $\langle \vec{S}_i \cdot \vec{S}_j \rangle$ corresponds to the generating function of walks between $i$ and $j$. In the high-temperature expansion, graphical representations of the correlation function involve both lines and closed loops. Each closed loop contributes a factor of $n$ due to the trace over internal indices. As $n \to 0$, such contributions vanish, leaving  a single open, non-self-intersecting path. These surviving configurations correspond to self-avoiding walks (SAWs), thereby establishing a mapping between the $O(n \to 0)$ field theory and the statistical ensemble of SAWs. Since the $n \to 0$ limit suppresses loops while retaining jump statistics, see Fig.\,\ref{fig:On_SAW_mapping}, the generalization of de Gennes' argument to the LR $O(n \to 0)$ model generates self-avoiding walks with LR jumps, referred to as self-avoiding L\'evy flights (L\'evy-SAW)~\cite{slade2018critical, slade2019self}.

\paragraph{SAW critical exponents:} The SAW serves as one of the simplest yet nontrivial critical systems to test and validate the field-theoretic prediction of universality and critical phenomena. The SAW critical exponents $\nu$ and $\gamma$ are defined as\,\cite{grassberger1985critical}:
\begin{align}
    \begin{split}
    \langle \log R_N \rangle &\sim \nu_{\rm{LR}} \log N,\\
    c_N &\sim \mu^N\, N^{\gamma-1},
    \end{split}
    \label{eq:nu_mu_gamma_def}
\end{align}
where $R_N$ and $c_N$ are respectively the \textit{end-to-end distance} and the number of $N$-step SAWs ($N \to \infty)$\,\cite{nu_SR_def}. The \textit{connective constant} is denoted by $\mu$. The exponent $\nu_{\rm{LR}}$ is generally referred to as the correlation-length exponent. 

The critical exponent $\gamma$, which is of our central interest, provides an understanding of phase transitions in SAWs as follows. Associating a weight factor $z$ with each SAW step, the two-point function becomes
\begin{align}
    G (R, z) = \sum_{N=0}^{\infty} c_N (R) z^N,
    \label{Eq:SAW_corr}
\end{align}
where $c_N(R)$ is the number of $N$-step SAWs that end at a distance $R$ from the origin. The susceptibility $\chi(z)$ is defined by the generating function\,\cite{slade2019self}
\begin{align}
    \chi(z)  =\sum_{R} G (R, z) = \sum_{N=0}^{\infty} c_{N} z^N,
    \label{Eq:SAW_susc}
\end{align}
where $c_N \coloneqq \sum_R c_N(R)$. Since $\chi(z)$ is a power series with the coefficients satisfying $c_N^{1/N} \to \mu$ in the limit $N \to \infty$, it has a radius of convergence $z_c = \mu^{-1}$, which is a nonuniversal parameter that depends on the underlying lattice and on $\sigma$.  
Then $\chi(z)$ is expected to diverge as
\begin{align}
    \chi(z) \sim (z_c -z)^{-\gamma} \quad \text{for} \quad z \nearrow z_c.
    \label{Eq:SAW_susc_crit}
\end{align}
The SAW thus exhibits a phase transition for all $\sigma > 0$ and $d \geq 1$, including in $d=2$ which for discrete symmetry is not constrained by the Mermin-Wagner theorem. 

The analogy between SAW and spin systems allows the computation of the SAW critical exponents using standard field-theoretic tools, e.g., $\epsilon$ expansion. In this context, the upper critical dimension of L\'evy-SAW is given by $d_\text{crit} = 2\sigma$ and the critical exponents for $d>d_\text{crit}$ simplify to $\nu = 1/\sigma,\,\text{and}\, \gamma = 1.$
Moreover, assuming the validity of the hyperscaling relation $\gamma = (2 - \eta) \nu$ , the LR-SR crossover for 1D L\'evy-SAWs reads,
\begin{equation}
\gamma =
\begin{cases}
\sigma \,\nu_{\rm{LR}}, & \sigma < \sigma^{*}, \\
\nu_{\rm{LR}}, & \sigma > \sigma^{*}.
\end{cases}
\label{eqn:L\'evy-SAW-exp-reln}
\end{equation}
Note that $\nu_{\rm{LR}}$ extracted from Eq.\,\eqref{eq:nu_mu_gamma_def} is identical to the usual $\nu_{\rm{SR}}$ in the SR scaling regime\,\cite{nu_SR_def}. In 1D SR SAWs, exact enumeration yields $\nu_{\rm{SR}} =1$ and $c_N=\text{const}$, implying $\gamma=1$. The hyper-scaling relation (whose validity will be discussed later) then yields $\eta = 1$, which, according to Sak's criterion, places the crossover at $\sigma^{*} = 1$\,\cite{madras2013self}.

This contradicts the scaling argument introduced by Flory, which  predicts $\nu_\text{LR}=3/(1+\sigma)$ for L\'evy SAWs in $d=1$ and $\nu_{\rm{SR}}=3/(2+d)$ in the SR case, seemingly implying a continuous crossover $\nu_\text{LR} = \nu_\text{SR}$ at $\sigma=2$ \cite{de1979scaling, halley1985node, grassberger1985critical}. The Flory picture is supported by Ref.\,\cite{grassberger1985critical}, whose MC study of L\'evy-SAW is not in agreement with the Sak's prediction in Eq.~\eqref{eqn:L\'evy-SAW-exp-reln}; instead, the exponent $\nu_{\rm{LR}}$ is shown to be smooth at $\sigma^{*} = 1$ and monotonically decreases with $\sigma$ up to $\sigma = 2$. Again, these contradictory results serve as evidence of the numerical difficulty of accessing the LR-SR crossover and extracting reliable estimates for the critical exponents.

In order to resolve this controversy we place particular focus on the LR-SR crossover region of L\'evy-SAWs in $d=1$. Our numerical results shed new light on the longstanding discrepancy between the field-theoretic predictions for $O(n)$ models and Flory-type scaling arguments,  going beyond previous numerical simulations and establishing a more complete understanding of the LR–SR crossover in low dimensions.

\paragraph*{Numerical simulation:}
In L\'evy flights, the probability for a step having a length $>r$ is assumed to decrease as
\begin{equation}
    P( l >r) \sim r^{-\sigma}, \quad \text{for } 0 < \sigma < 2.
\end{equation}
Let us proceed to simulate L\'evy-SAW on a 1D infinite lattice. For a given $\sigma$, in MC simulations, the probability of a walker making a step of length $r$ either to the left or right is chosen as
$P(r) = \left [ r^{-\sigma} - (r+1)^{-\sigma}\right]/ 2$\,\cite{grassberger1985critical}. While short walks are affected by the precise form of $P(r)$, the critical scaling is determined by long jumps with $P(r)\sim r^{-(1+\sigma)}$, and critical exponents are reliably determined only for large walk lengths. In simulations, the walk starts at the origin. At each step, the walker draws a random step-length $r$ from the given distribution, and attempts to hop to a site at distance $r$ to the left or right of its current position. If the new position has not been visited before, the walker updates its position, draws another $r$ and repeats the process. Otherwise, the walk terminates.

\paragraph*{L\'evy-SAW with branching:}
Within the $\sigma$ range of interest, the typical lengths $N=10\cdots30$ of successful walks are too short to reach the scaling regime and determine the exponents reliably. To overcome this limitation, we introduce an enrichment algorithm, the \textit{branching SAW,} where at each node the walker branches out and makes $z>1$ attempts for the next step (convergence requires $z<z_c$). According to Eqs.\,(\ref{Eq:SAW_susc},\ref{Eq:SAW_susc_crit}), as criticality $z\to z_c$ is approached arbitrarily long walks are sampled and one has to cut off at a maximum $N$, with runtime scaling as $N^{1+\gamma}$. In practice, at each node, the walker always performs the first step, and attempts the second step with a probability $z-1$. The branching does not alter the critical scaling but provides better statistics for the reliable estimation of the critical exponents\,\cite{sokal1994monte, binder-book}.

\paragraph*{Results:} 
We first examine the behavior of the anomalous dimension $\eta$ as extracted from MC simulations. In simulations, we set the branching factor $z(\sigma)$ close to $z_c(\sigma)$ to generate SAW data up to $2\times 10^3$ (in the crossover $10^4$) steps such that it ensures the longest walks have statistics of at least $10^7$ samples, while the shortest walks have at least $10^9$ samples. 

\begin{figure}[htbp!]
\centering
\includegraphics[width=0.95\linewidth]{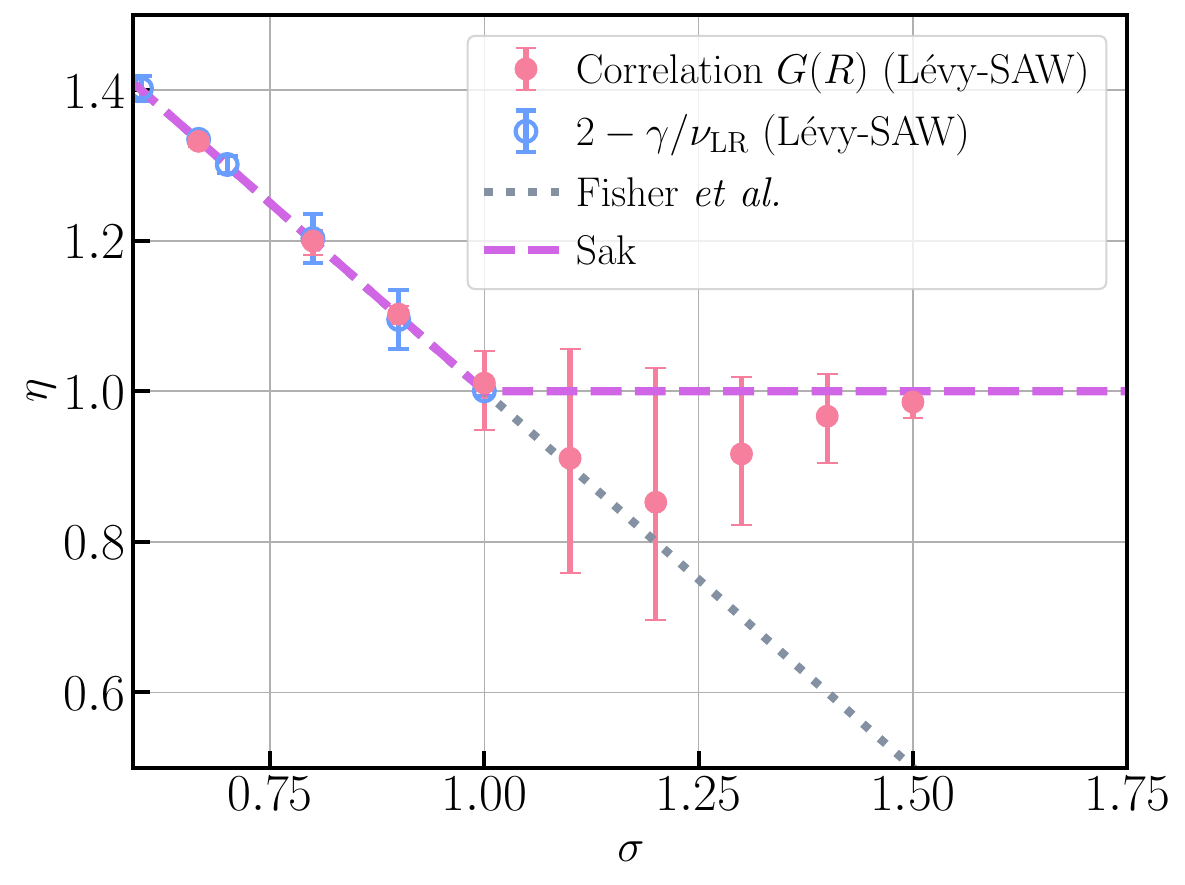}
    \caption{Critical exponent $\eta$ vs.\ $\sigma$ across the LR-SR crossover. The data (filled pink circles) interpolate between Fisher’s prediction $\eta = 2 - \sigma$ and the SR value $\eta_{\rm{SR}} = 1$, and are in agreement with Sak’s scenario. The empty blue circles correspond to $2 - \gamma/\nu_{\rm{LR}}$, validating the hyperscaling relation.}
    \label{fig:eta_SAW}
\end{figure}

To determine $\eta (d,\sigma)$, we compute the two-point correlation function \eqref{Eq:SAW_corr} at $z \to z_c$ and extract the exponent by fitting to the expected power-law, $G(R) \sim R^{1-\eta}$\,\cite{SM}. Figure\,\ref{fig:eta_SAW} shows the behavior of $\eta$ with $\sigma$ across the LR-SR crossover. For $\sigma < 1$, the results are in excellent agreement with the conjecture $\eta (d, \sigma) = 2 - \sigma$. At $\sigma = 1$, $G(R)$ becomes almost flat showing convergence towards the SR value $\eta_{\rm{SR}} =\eta (d=1, \sigma \to \infty) = 1$, with our estimate $\eta = 1.011_{-0.062}^{+0.042}$. A careful analysis \,\cite{SM} reveals a weak logarithmic correction at this point, as expected by field theoretic arguments\,\cite{angelini2014relations, brezin2014crossover, defenu2015fixed}.

For $\sigma > 1$, the exponent remains close to SR value, $\eta_{\rm{SR}} = 1$. At $\sigma = 1.1$ and $1.2$, large error bars include both Fisher and Sak predictions, although at $\sigma = 1.2$, our MC estimate clearly deviates from Fisher toward Sak. This trend is more evident at $\sigma=1.3$ and $1.4$, and by $\sigma = 1.5$, $\eta$ aligns with $\eta_{\rm{SR}}$, which supports the Sak scenario over Fisher’s. Large uncertainties appear at $\sigma \gtrsim 1$, where our maximum walk lengths are beyond LR scaling but have not yet reached the SR scaling regime. Our data are thus in agreement with Sak’s prediction, continuously interpolating between the LR and SR regimes around $\sigma^* \approx 1$. The absence of the expected cusp at $\sigma=1$ is attributed to the slow convergence to the asymptotic scaling form (see below). Moreover, the small deviations from $\eta_{\rm{SR}}$ near the crossover are insufficient to justify a transition at $\sigma = 2$, as implied by Fisher's original criterion~\cite{fisher1972critical} and previous MC results~\cite{grassberger1985critical}. Overall, our results corroborate Sak’s scenario and strongly suggest the crossover near $\sigma^* = 1$.

\begin{figure}[tbp!]
\centering
\includegraphics[width=0.95\linewidth]{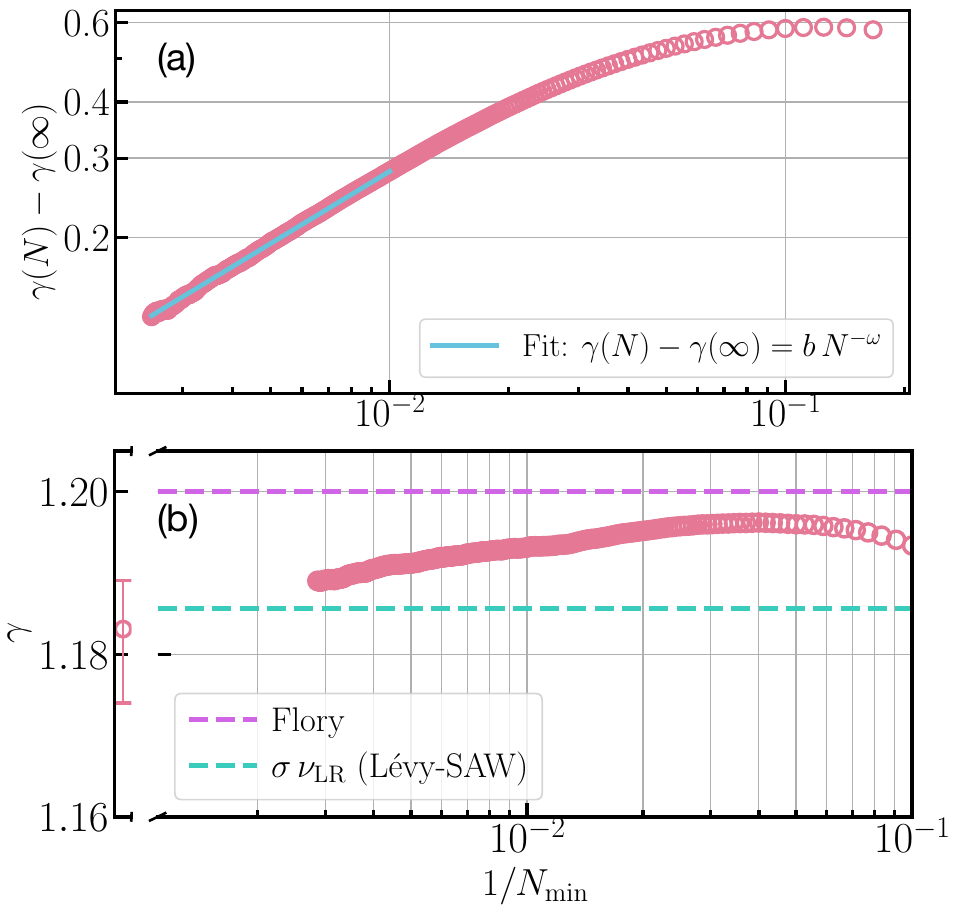}
\caption{Scaling of the SAW exponent $\gamma$ with inverse walk length in the SR, panel (a), and LR regimes, panel (b). At $\sigma = 1.5$ in the SR regime, $\gamma$ approaches the asymptotic value $\gamma(\infty) = 1$, with a finite-size correction exponent $\omega = 0.533(1) \approx \sigma - \sigma^{*}$. At $\sigma = 2/3$ in the LR regime, our (asymptotic) MC estimate is $\gamma = 1.183^{+0.005}_{-0.009}$. The dashed violet line shows the Flory prediction, while the dashed green line represents the prediction $\gamma = \sigma \nu_{\rm{LR}}$ based on our measured $\nu_{\rm{LR}}$ \,\cite{SM}.}
\label{fig:gamma_scaling_LR_SAW}
\end{figure}

\paragraph*{Critical exponents:}
To provide clear evidence about the LR-SR crossover around $\sigma^* = 1$, we focus our discussion on the behavior of critical exponent $\gamma$ with walk length. Sizable corrections to scaling mean that the apparent exponent extracted at length $N$ converges only slowly to the asymptotic $N\to\infty$ value, $\gamma(N) = \gamma(\infty)+bN^{-\omega}$ \cite{SM}, see Fig.~\ref{fig:gamma_scaling_LR_SAW}. All the critical exponents presented in this Letter are estimated from the MC data by analyzing $c_N$ for $\gamma$ and the average logarithm, $\langle \log R_N \rangle$, for $\nu$ \,\cite{SM}.

Figure~\ref{fig:gamma_scaling_LR_SAW}(a) shows the scaling behavior of $\gamma$ with inverse walk-length $N_{\rm{min}}$\,\cite{SM} at $\sigma = 1.5$. Such a choice within the intermediate regime $\sigma \in [1,2]$ minimizes boundary effects and associated numerical uncertainties. The exponent is expected to approach asymptotically $\gamma(N\to\infty) = 1$, according to the Sak prediction, but in contrast to the Flory prediction. Field theory predicts a finite-size correction to the scaling $\sim N^{-(\sigma - \sigma^*)}$\,\cite{SM}. With $\gamma(\infty) = 1$, a fitting of the data using $\gamma (N) - \gamma(\infty) = b N^{-\omega}$, yields $\omega = 0.533(1) \approx \sigma - \sigma^{*}$. This provides strong evidence in favor of the Sak scenario and convincingly excludes the crossover at $\sigma=2$. Such a precise verification is made possible by our branching algorithm, which substantially improves the statistics of long walks. 

We now turn our attention to the LR regime. As a representative case, consider $\sigma=2/3$, which lies well between the MF-LR and LR-SR crossover boundaries, see Fig.\,\ref{fig:gamma_scaling_LR_SAW} (b). The exponent exhibits a non-monotonic behavior with the inverse walk length. A detailed analysis incorporating corrections to scaling yields an asymptotic value $\gamma = 1.183^{+0.005}_{-0.009}$, which deviates from the Flory predictions (violet dashed line) by $ \approx 1.4\%$. In fact, both our finite-walk length data and extrapolated asymptotic value lie slightly below the Flory estimate.  

The results in Fig.\,\ref{fig:gamma_scaling_LR_SAW}(b) can be used to test the validity of the hyperscaling relation in Eq.~\eqref{eqn:L\'evy-SAW-exp-reln}. The dashed green line marks the hyperscaling prediction based on our MC estimate of $\sigma\, \nu_{\rm{LR}}$ \,\cite{SM}, and it agrees within error bars with our MC estimate for $\gamma$, thereby supporting Eq.~\eqref{eqn:L\'evy-SAW-exp-reln}. A more comprehensive validation across the entire LR regime is presented in Fig.\,\ref{fig:eta_SAW} (empty blue circles), where $ 2-\gamma / \nu_{\rm{LR}}$ is plotted up to $\sigma^* = 1$. This quantity closely matches  $\eta$ within error bars. Note that in the SR regime, the hyperscaling relation always holds with $\gamma =\nu_{\rm{SR}}=1$.

\begin{figure}[tbp!]
\centering
\includegraphics[width=0.95\linewidth]{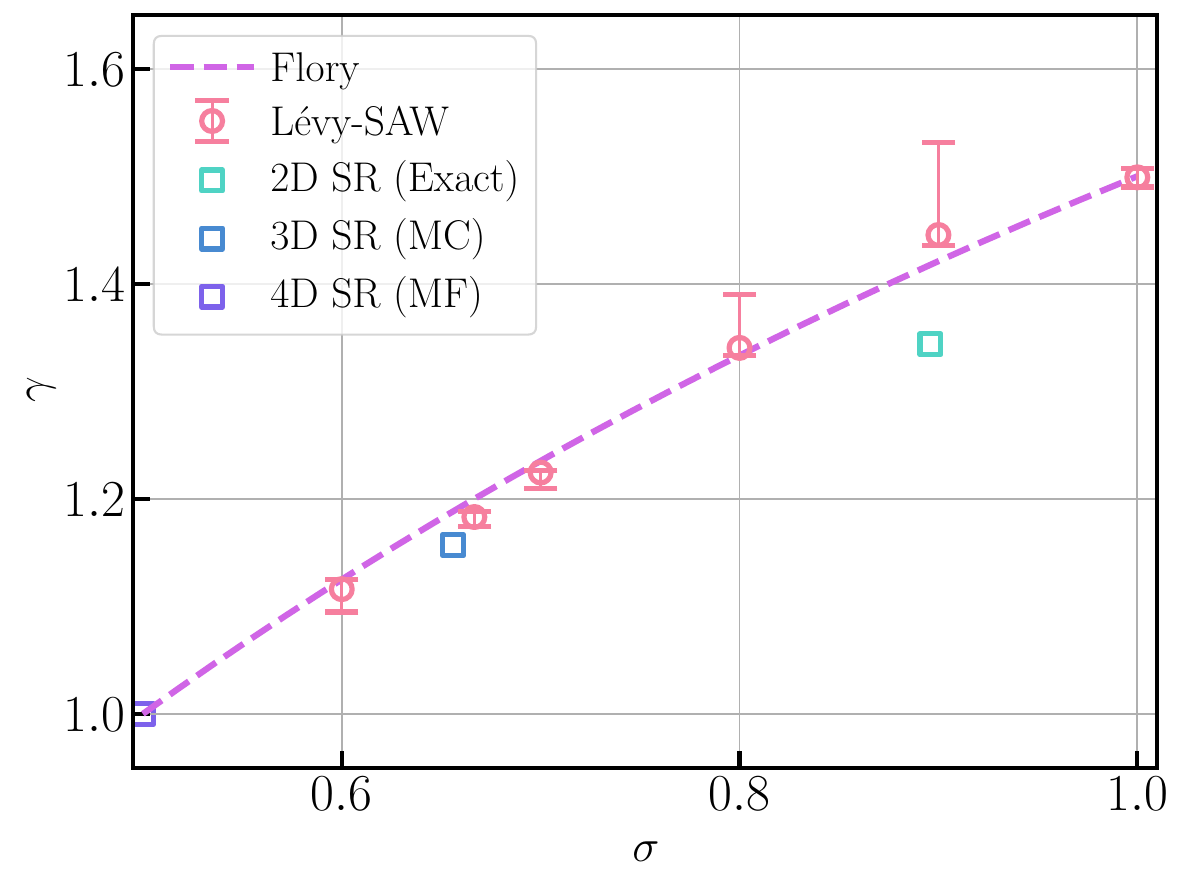}
\caption{Critical exponent $\gamma$ in the LR regime. Our MC results (red circles) are close to the Flory prediction (dashed violet line) in the entire LR regime, which extends up to $\sigma^*=1$. Individual SR results (squares) are compared at matching effective dimension $D_\text{eff}$.}
\label{fig:nu_gamma_SAW}
\end{figure}

We proceed to investigate the behavior of the exponent $\gamma$ at various $\sigma$ over the entire LR regime, see Fig.~\ref{fig:nu_gamma_SAW}. Several key observations are in order. First, $\gamma$ varies monotonically with increasing $\sigma$ up to $\sigma^{*} = 1$. The deviation from the Flory prediction remains rather small in the entire $\sigma$ range, despite being clearly visible in the scaling analysis of Fig.\,\ref{fig:gamma_scaling_LR_SAW}. 

Second, an important aspect of our analysis is the consideration of corrections to scaling exponents, which are universal\,\cite{sokal1994monte} and depend on the distance from the LR-SR boundary. In particular, logarithmic corrections are expected at the boundary, and are indeed observed for $\eta$ at $\sigma^{*} = 1$\,\cite{SM}. A semi-quantitative understanding of universal behavior in the LR regime can be obtained by the effective dimension approach\,\cite{solfanelli2024universality}, where thermodynamic exponents at a given $\sigma$ correspond to those of the SR model in an effective dimension $D_{\rm{eff}} = (2 - \eta(D_{\rm{eff}}, \sigma \to \infty))d/\sigma$. It is thus worthwhile to compare SR results for $\gamma$ at matching $D_{\rm{eff}}$ with LR Flory. For instance, $D_{\rm{eff}}=3$ and $2$ are reached at $\sigma \approx 0.656$, and $ 0.896$, with corresponding Flory estimates $\gamma = 1.188$ and $1.418$, showing $\approx 2.7\%$ and $5.5\%$ deviation from the $3$D MC: $\gamma = 1.156\cdots$ (blue square) and the exact 2D field-theoretic result:  $\gamma = 43/32$ (green square), respectively.

As expected\,\cite{defenu2015fixed, angelini2014relations, behan2017scaling}, the effective dimension prediction $\gamma(d,\sigma) = \gamma(D_{\rm{eff}}, \sigma \to \infty)$ holds only approximately for LR systems. The discrepancy is expected to grow with $\eta_{\rm{SR}}$, reflecting stronger correlations in the critical theory\,\cite{defenu2015fixed}. Consequently, this approach is not as accurate in $d=1$ with its large $\eta_{\rm{SR}}$ compared to the Ising case in $d=2$\,\cite{solfanelli2024universality}.

To summarize, we numerically demonstrate, by studying the exponent $\eta$, that the LR-SR critical behavior in a $1$D LR $O(n \to 0)$ model follows Sak's picture with a crossover at $\sigma^* =1$, see Fig.\,\ref{fig:eta_SAW}. The predictions by Fisher and Flory are further ruled out by examining the scaling behavior of the critical exponent $\gamma$, see Figs.\,\ref{fig:gamma_scaling_LR_SAW}, \ref{fig:nu_gamma_SAW}. In the LR regime, we find that the exponent remains close to, but deviates from, the Flory prediction, except at the LR-SR crossover point.  In the SR regime $\sigma > 1$, $\gamma$ asymptotically approaches $\gamma=1$.
Thus, our findings differ from previous claims in Ref.\,\cite{grassberger1985critical}, where a smooth dependence of $\gamma$ on $\sigma$ up to $\sigma = 2$ was claimed. With our improved algorithm, we clarify these strong finite-size effects and unveil the true LR–SR crossover, consistent with Sak’s scenario. Furthermore, our findings highlight that effective spatial dimension alone does not dictate universality when LR interactions and nontrivial anomalous dimensions are present.

This work suggests a natural extension toward exploring universality in the presence of disorder. In particular, generalizing to sparse random graphs with LR links \,\cite{millan2021complex, sarkar2024universality, bighin2024universal}, which lack translational symmetry, offers a rich playground to explore the interplay between geometry, connectivity, and criticality, with potential relevance to quantum entanglement, information spreading, and emergent dynamics in complex networks\,\cite{king2025optimal}.

\begin{acknowledgments}
We acknowledge fruitful discussions with Giacomo Gori. This research was funded by the Swiss National Science Foundation (SNSF) grant numbers 200021--207537 and 200021--236722, by the Deutsche Forschungsgemeinschaft (DFG, German Research Foundation) under Germany's Excellence Strategy EXC2181/1-390900948 (the Heidelberg STRUCTURES Excellence Cluster) and the Swiss State Secretariat for Education, Research and Innovation (SERI). M.S. also acknowledges support by the state of Baden-Württemberg through bwHPC cluster.
\end{acknowledgments}



\bibliographystyle{apsrev4-2.bst}
\bibliography{References_mrinal}

\clearpage
\onecolumngrid  
\appendix
\begin{center}
	\textbf{\large Supplemental Material for ``Long range to short range crossover in one dimension''}\\[5pt]
    Mrinal Sarkar$^1$, Nicolò Defenu$^2$, and Tilman Enss$^1$ \\[3pt]
    \textit{$^1$Institut für Theoretische Physik, Universität Heidelberg, 69120 Heidelberg, Germany} \\[3pt]
    \textit{$^2$Institut für Theoretische Physik, ETH Zürich, Wolfgang-Pauli-Str.27, 8093 Zürich, Switzerland
} \\[3pt]
( Dated:\,\today)
\end{center}
\bigskip

\renewcommand*{\citenumfont}[1]{S#1}
\renewcommand*{\bibnumfmt}[1]{[S#1]}
\renewcommand{\theequation}{S\arabic{equation}}
\renewcommand{\thetable}{S\arabic{table}}
\renewcommand{\thefigure}{S\arabic{figure}}
\setcounter{equation}{0}
\setcounter{table}{0}
\setcounter{figure}{0}

\section*{Table of Contents}
\begin{enumerate}
    \item Extraction of the anomalous dimension $\eta$\dotfill \pageref{supp-sec:eta-extraction}
    \item Extraction of SAW critical exponents \dotfill 
    \pageref{supp-sec:exp-extraction}
    \item Finite size correction to SAW critical exponents \dotfill \pageref{supp-sec:FS-correction}
    \item Behavior of the SAW critical exponent $\nu$ \dotfill \pageref{supp-sec:SAW-exponent-nu}  
\end{enumerate}

\bigskip
\section{Extraction of the anomalous dimension $\eta$}
\label{supp-sec:eta-extraction}
To extract the exponent $\eta$ for a given $\sigma$, we proceed as follows:

First, in the simulations, we set a value of $z$ close to $z_c$ (within 1\% of $z_c$) and record the data $c_N(R)$ up to a given $N = 10^4$ and $R = 10^4$.
Next, the procedure involves two steps:

1. \textbf{Extrapolation to criticality:}  
We first extrapolate $c_N(R)$ from the value at $z$ to the critical value at $z = z_c$, as follows:
\begin{equation}
c_{N}^{\text{crit}}(R) = c_N(R) \cdot \left (\frac{z_c}{z} \right)^N.
\end{equation}

2. \textbf{Tail correction and computation of $G(R)$:}  
We then extract the tail exponent from the asymptotic scaling of $c_N^{\text{crit}}(R)$ at large $N$ 
\begin{equation}
C_N(R) \sim N^{-{\rm{exponent}}},
\end{equation}
which is used to correct the tail of the $c_N(R)$ data. Using the corrected data, we compute $G(R)$ according to Eq.~(2) in the main text.
Due to finite-size limitations in simulations, only data up to $N_{\text{stop}}$ is available with good statistics, so we estimate the tail beyond $N_{\text{stop}}$ using this power-law:
\begin{equation}
G(R) = G(R, z_c) \approx \sum_{N=1}^{N_{\text{stop}}} c_N^{\text{crit}}(R)
+ \frac{c_{N_{\text{stop}}}^{\text{crit}}(R) \cdot N_{\text{stop}}}{\rm{exponent} - 1}.
\end{equation}

Next, we extract the exponent $\eta$ from the relation $G(R) \sim R^{1- \eta}$ using the following formula:
\begin{equation}
\eta(R_{\rm{start}}, R) = \frac{R \cdot \frac{1}{G(R)} - R_{\text{start}} \cdot \frac{1}{G(R_{\text{start}})}}{\displaystyle\sum_{r = R_{\text{start}} + 1}^{R-1} \frac{1}{G(r)} + \frac{1}{2} \left( \frac{1}{ G(R_{\rm{start}})} +\frac{1}{ G(R)}\right)}.
\label{eq:supp_eta_G_formula}
\end{equation}

To improve the robustness of the extraction, we compare two methods. One is based directly on $1/G(R)$, and the other uses $R/G(R)$.  The latter grows more strongly with $R$ (as $R^{\eta}$ instead of $R^{\eta - 1}$), and hence gives relatively more weight to longer walks. While both methods yield consistent average value for $\eta$, the method using $R/G(R)$ yields a large error bar. We adopt the latter one as it provides a more conservative uncertainty estimates.

For both methods, we compute $\eta(R_{\text{start}}, R)$ for different values of $R_{\text{start}}$, and study the asymptotic behavior of the resulting $\eta$ curves. The curves corresponding to different $R_{\text{start}}$ values typically converge beyond a certain $R$. The final value of $\eta$ is estimated by averaging $\eta(R)$ over the range where convergence occurs.

To estimate the uncertainty in the exponent, we take the maximum and minimum values of $\eta(R)$ in the same interval, which define the upper and lower bounds. These bounds are used to determine the error bars shown in Fig.~\ref{fig:eta_SAW}.

\begin{figure*}[htbp!]
\centering
\includegraphics[width=0.45\linewidth]{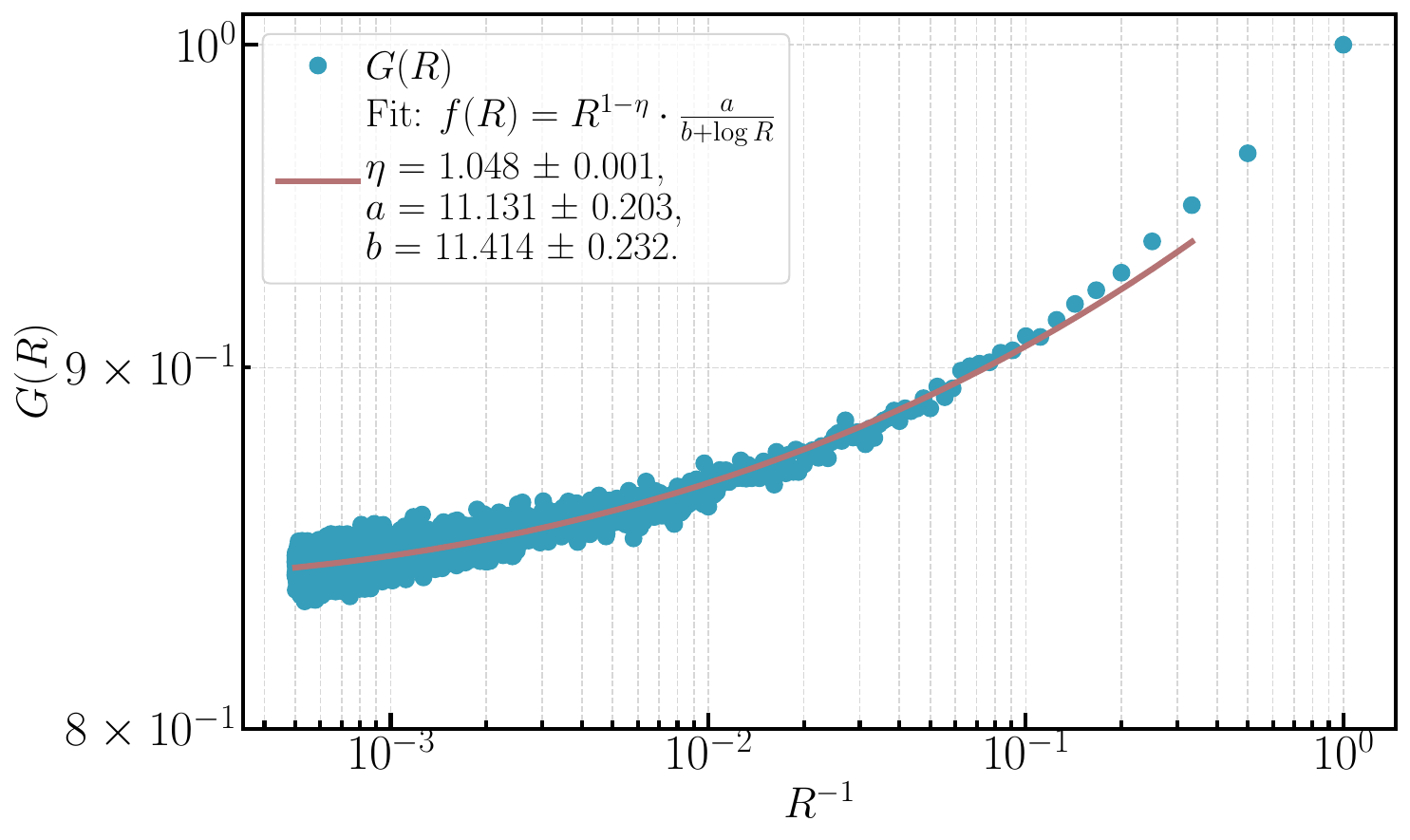}
\caption{Logarithmic correction to scaling of the two-point correlation $G(R)$, evaluated at $z_c$, for $\sigma=1$.}
\label{fig-supp:log_correction_sigma1}
\end{figure*}

Before applying Eq.\,\eqref{eq:supp_eta_G_formula}, we first examine the behavior of $G(R)$ versus $R^{-1}$ curves at various values of the LR decay parameter $\sigma$. At $\sigma = 1$, we indeed observe the presence of a weak logarithmic corrections in $G(R)$. One would naturally expect such corrections to appear only at large $R$. Fitting our MC data with the form
\begin{equation}
G(R) = R^{1-\eta} \cdot \frac{a}{b + \log(R)},
\end{equation}
we obtain the parameters $\eta = 1.048(1)$, $a = 11.1(2)$, and $b = 11.4(2)$, see Fig.\,\ref{fig-supp:log_correction_sigma1}. These results reveal the subtle nature of the observed logarithmic corrections.

\section{Extraction of SAW critical exponents}
\label{supp-sec:exp-extraction}
All the critical exponents presented in this letter are estimated from the MC data by analyzing the number of walks $c_N$, for $\gamma$, and the average logarithm of end-to-end distance, $\langle \log R_N \rangle$, for $\nu_\text{LR}$. Specifically, both $\gamma$ and $z_c$ are estimated using the relation
\begin{equation}
\log\left(\frac{c_N}{c_{N-1}}\right) = -\log\left(\frac{z_c}{z}\right) + \frac{\gamma - 1}{N},
\end{equation}
and the exponent $\nu$ is estimated following the scaling relation
\begin{align}
    \langle \log R_N \rangle &\sim \nu\, \log N.
\end{align}

To extract these exponents, the data is divided into multiple windows, where each starting at walk length $N_{\rm{min}} = i$, say, and extending up to $4 i$, covering the entire dataset. Using different window sizes in step length helps reduce statistical fluctuations and improve accuracy. Within each window, the exponents are extracted by fitting the appropriate scaling functions to the weighted data. Finally, the estimated exponents are plotted against the inverse of the starting walk length in each window to examine their scaling behavior and finally determine their values in the asymptotic limit with asymmetric error bars.

\section{Finite size correction to SAW critical exponents}
\label{supp-sec:FS-correction}
One may justify the appearance of finite-size scaling with $(\sigma-\sigma*)$ as follows. Defenu {\it{et. al.}}\,\cite{defenu2015fixed} studied the $O(n)$ model with LR interaction $\sim r^{-(d+\sigma)}$, using functional renormalization group (fRG) method, with a focus on $n=1$ and $d>=2$. To understand the competition between short-range (SR) and long-range (LR) interactions, an effective action containing both LR and SR terms was introduced:
\begin{align}
\Gamma_k[\phi] = \int d^{d}x \left\{ Z_{\sigma} \partial_{\mu}^{\frac{\sigma}{2}} \phi_{i} \partial_{\mu}^{\frac{\sigma}{2}} \phi_{i} + Z_2 \partial_{\mu} \phi_{i} \partial_{\mu} \phi_{i} + U_k(\rho) \right\}.
\label{eq:fRG-action-ansatz}
\end{align}
where the summation over repeated indices is assumed, $\rho = \frac{1}{2} \phi_i \phi_i$, and $\phi_i$ is the $i$-th component of $\bf{\phi}$. The first term in the action corresponds to the LR part, where the inverse propagator in Fourier space behaves like $q^{\sigma}$, while the second term corresponds to the SR case, where the inverse propagator behaves as $q^2$. The term $Z_k$ is the wave-function renormalization of the model, which is approximated to be field-independent and $U_k(\rho)$ is the effective potential. Defining a generalized Litim cutoff,
\begin{align}
    R_k(q) = Z_{\sigma} (k^{\sigma} - q^{\sigma}) \theta(k^{\sigma} - q^{\sigma}) + Z_2 (k^2 - q^2) \theta(k^2 - q^2),
    \label{eq:fRG-Litim_cutoff}
\end{align} the flow equations for the renormalized dimensionless couplings are derived in the SR regime. These are given by\,\cite{defenu2015fixed}
\begin{align}
\partial_t \bar{J}_\sigma &= (\sigma - 2) \bar{J}_\sigma + \eta_2 \bar{J}_\sigma, \label{eq:fRG-flow_J} \\
\eta_2 &= \frac{(2 + \sigma \bar{J}_\sigma)^2 \bar{\rho}_0 U_k''(\bar{\rho}_0)^2}
{(1 + \bar{J}_\sigma)^2 [1 + \bar{J}_\sigma + 2 \bar{\rho}_0 U_k''(\bar{\rho}_0)]^2}, \label{eq:fRG-flow_eta} \\
\partial_t \bar{U}_k(\bar{\rho}) &= -d \bar{U}_k(\bar{\rho}) + (d - 2 + \eta_2) \bar{\rho} \bar{U}_k'(\bar{\rho}) + (n - 1) \frac{1 - \eta_2}{d+2 + \frac{\sigma}{2} \bar{J}_\sigma} \frac{1}{1 + \bar{J}_\sigma + \bar{U}_k'(\bar{\rho})} \notag \\
&\quad + \frac{1 - \eta_2}{d+2 + \frac{\sigma}{2} \bar{J}_\sigma} \frac{1}{1 + \bar{J}_\sigma + \bar{U}_k'(\bar{\rho}) + 2 \bar{\rho} \bar{U}_k''(\bar{\rho})}. \label{eq:fRG-flow_potential}
\end{align}
Here $t=\log(k/k_0)$ is the RG time, with $k_0$ being the ultraviolet scale, $\eta_2 = - {Z_2}^{-1} \partial_t Z_2$ is the anomalous dimension, and $J_\sigma = {Z_\sigma}/{Z_2}$ is the renormalized LR coupling.

Assuming that Eqs.\,\eqref{eq:fRG-flow_J}, \eqref{eq:fRG-flow_eta}, \eqref{eq:fRG-flow_potential} hold for $n=0$ and $d=1$, and identifying $\sigma^{*} = 2 - \eta_2$, one obtains from the evolution equation for $J_\sigma$ in the regime $ \sigma > \sigma^* $,
\begin{align}
    \partial_t \bar{J}_\sigma = (\sigma - \sigma^*) \bar{J}_\sigma,
\end{align}
yielding
\begin{align}
    \bar{J}_\sigma (N) = \bar{J}_\sigma (0)N^{(\sigma - \sigma^*)} ,
\end{align}

This shows that $ \Delta = \sigma - \sigma^*$ is the scaling dimension associated with $J_\sigma$. Since $J_\sigma$ is the ratio of the LR and SR couplings, this indicates how the system flows towards either a LR or SR fixed point under renormalization. In the SR regime $(\sigma > \sigma^*)$, $J_\sigma$ is irrelevant variable, implying that SR fixed point is stable. However, in the LR regime $(\sigma < \sigma^*)$, $J_\sigma$ grows under RG flow. At $\sigma=\sigma^*$, LR fixed point appears and controls the critical behavior in the region $1/2 < \sigma < \sigma^*$\,\cite{defenu2015fixed}.

The evolution of $\bar{J}_\sigma$, being an irrelevant operator in SR regime, influences the scaling of physical quantities such as the correlation length. To understand how this coupling contributes to the finite-size corrections to the correlation length exponent of SAW, recall that the correlation length $\xi$ in SAW corresponds to the end-to-end distance. By definition, in the thermodynamic limit, the correlation length scales as $\xi (\infty) = R_{N}(N  \to \infty) \sim N_{\infty}^{\nu (\infty)}$. Here $N_{\infty}$ represents very large walk lengths for which the condition $N\to \infty$ applies. In finite system with an $N$-step walk, $\xi (N) \sim N^{\nu (N)}$, where $\nu (N)$ is the effective correlation length exponent in a finite system. In the thermodynamic limit, $\xi (N)$ approaches $\xi (\infty)$, but finite-size corrections will modify the scaling. In general, one may write\,\cite{sokal1994monte,hughes1996random}
\begin{align}
    \xi(N) = N_{\infty}^{\nu(\infty)} \left( 1 + aN^{-\Delta} + O\left(N^{-(\Delta +1)}\right)\right),
\end{align}
where $a$ is some non-universal constant. For large $N$, this leads to
\begin{align}
    \nu (N) = \nu(\infty) +  b N^{-\Delta},\label{eq:nu_scaling_correction}
\end{align}
with $b = {a}/{\log N_{\infty}}$. 
The exponent $\Delta = \sigma-\sigma^*$ dictates how fast the system approaches pure SR behavior as $N\to \infty$. Note that for $\sigma \gtrsim \sigma^*$, the decay is very slow, implying that finite-size corrections are very strong even at large $N$.

In the LR regime ($\sigma < \sigma^*)$, $\bar{J}_\sigma$ is a relevant operator. Also, $Z_2$ is well defined in this case and is not brought to zero by the presence of a dominant LR term. However, due to the presence of residual SR contributions in finite systems, we expect subleading finite-size corrections of the same form as in Eq.\,\eqref{eq:nu_scaling_correction}, but with different coefficients.

We expect similar finite-size corrections to apply to all thermodynamic observables, and therefore anticipate similar correction to scaling for all the critical exponents including the susceptibility exponent $\gamma$.

\section{Behavior of the SAW critical exponent $\nu$ }
\label{supp-sec:SAW-exponent-nu}

\begin{figure}[htbp!]
\centering
\includegraphics[width=0.425\linewidth]{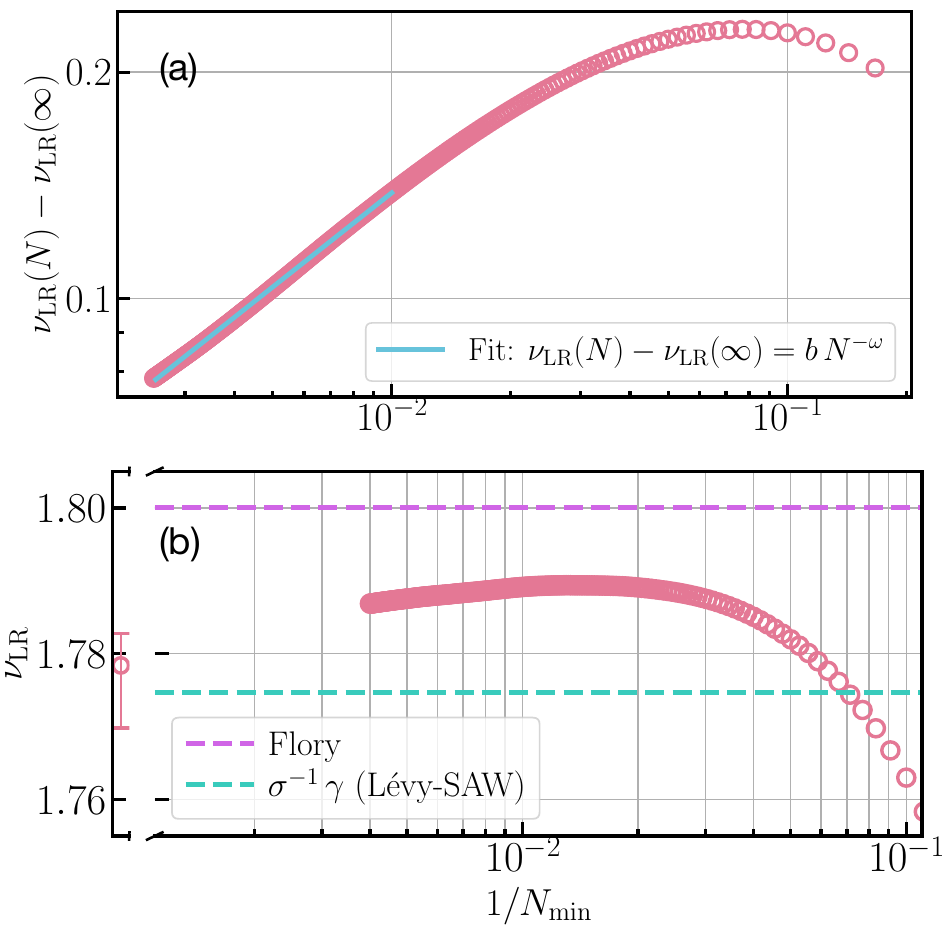}
\caption{Scaling of the SAW exponent $\nu_{\rm{LR}}$ with inverse walk length in the SR, panel (a), and LR regimes, panel (b). At $\sigma = 1.5$ (SR regime), $\nu$ approaches the asymptotic value $\nu_{\rm{LR}}(\infty) = 1$, with a finite-size correction exponent $\omega = 0.417(1)$ which is close to $\sigma - \sigma^{*}$. At $\sigma = 2/3$ in the LR regime, the dashed violet line shows the Flory prediction, while the dashed green line represents the hyperscaling prediction $\nu_{\rm{LR}} = \sigma^{-1} \gamma$ based on our MC estimate $\gamma$.}
\label{fig:app_nu_scaling}
\end{figure}

\begin{figure}[htbp!]
\centering
\includegraphics[width=0.425\linewidth]{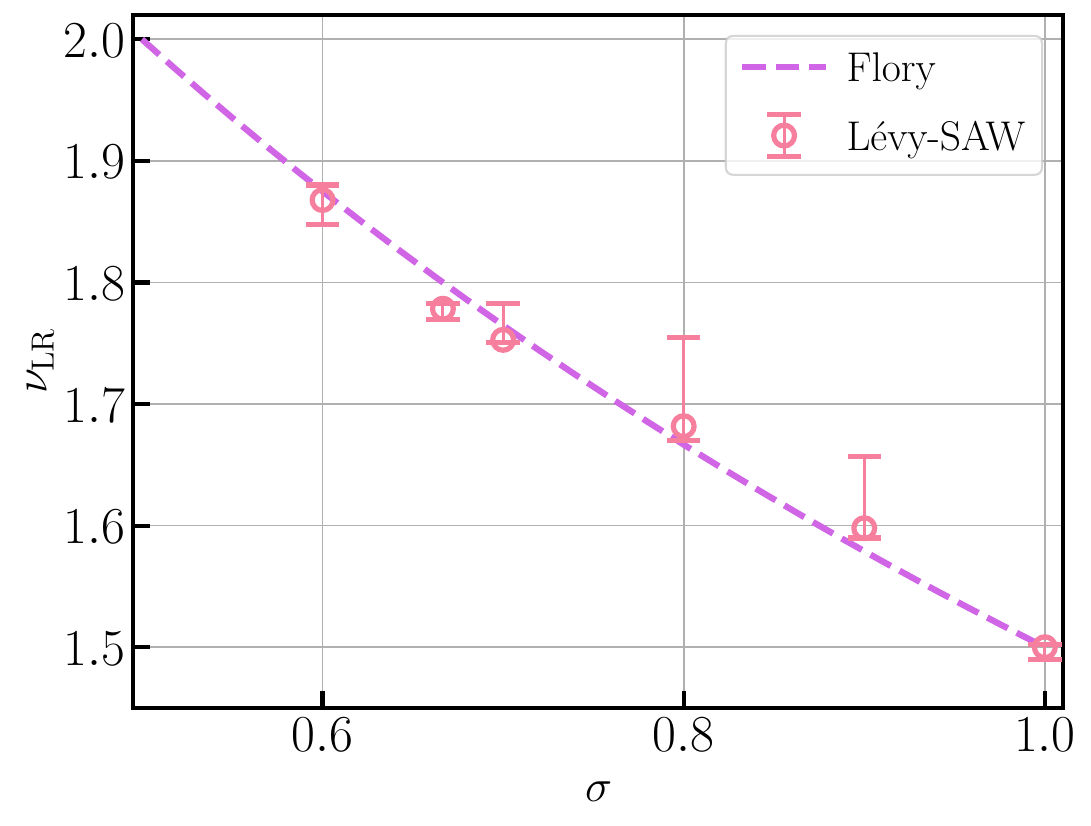}
\caption{Critical exponent $\nu$ vs. LR parameter $\sigma$ in the LR regime. Our MC results (red circles) are close to the Flory prediction (dashed violet line) in the entire LR regime, i.e., up to $\sigma^*=1$.}
\label{fig:app_nu_vs_sigma_SAW}
\end{figure}

Here we present additional evidence supporting the LR-SR crossover at $\sigma^* = 1$, based on the behavior of the SAW critical exponent, $\nu_{\rm{LR}}$, with walk-lengths, see Fig.~\ref{fig:app_nu_scaling}. These results have already been used in the main text to validate the hyperscaling relation in Fig.\ref{fig:eta_SAW}. Here we provide a complementary picture from the perspective of $\nu_{\rm{LR}}$.

Similar to Fig.\,\ref{fig:gamma_scaling_LR_SAW} in the main text, Fig.~\ref{fig:app_nu_scaling} (a) shows the scaling behavior with inverse walk-length of $\nu_{\rm{LR}}$ at $\sigma = 1.5$. The asymptotic value of the exponent is expected to approach $\nu_{\rm{LR}}(\infty) = 1$, consistent with Sak prediction and in contrast to the Flory prediction. With $\nu_{\rm{LR}}(\infty) = 1$, a fitting of the data using a power-law $\nu_{\rm{LR}} (N) - \nu_{\rm{LR}}(\infty) = b N^{-\omega}$ to incorporate finite-size correction to the scaling, yields $\omega = 0.417(1)$, which shows agreement with field theoretic prediction $\sigma - \sigma^{*}$, providing strong evidence in favor of the Sak scenario, and convincingly exclude the crossover at $\sigma=2$.

Now in the LR regime, at $\sigma=2/3$, the exponent exhibits a non-monotonic behavior with the inverse walk length. This is shown in Fig.\,\ref{fig:app_nu_scaling}(b). A detailed analysis incorporating corrections to scaling yields an asymptotic value $\nu_{\rm{LR}} = 1.778^{+0.004}_{-0.009}$, which deviates from the Flory predictions ($\nu_{\rm{LR}} = 1.8$; shown as a violet dashed line) by $ \approx 1.2\%$. In fact, both our finite-walk length data and its extrapolated asymptotic value lie slightly below the Flory estimate. To test the validity of the hyperscaling relation, Eq.~\eqref{eqn:L\'evy-SAW-exp-reln}, we show in Fig.\,\ref{fig:app_nu_scaling}(b) a dashed-line (green) corresponding to $\sigma^{-1} \gamma$, which lies within the error bars of the estimated $\nu_{\rm{LR}}$, thereby supporting Eq.~\eqref{eqn:L\'evy-SAW-exp-reln}. 

We then proceed to compute the behavior of the $\nu_{\rm{LR}}$ at various $\sigma$ values over the entire LR regime, as shown in Fig.~\ref{fig:app_nu_vs_sigma_SAW}. We observe, here as well, that the exponent $\nu_{\rm{LR}}$ remains close to but consistently deviates from the Flory prediction. 

\end{document}